\documentstyle[epsfig,twocolumn,seceq]{asr2k}

\title{ Soft Phonon Anomalies in Relaxor Ferroelectrics }

\def\runtitle{ Soft Phonon Anomalies in Relaxor Ferroelectrics }

\author{ Gen {\sc SHIRANE} and Peter {\sc M.\ GEHRING}$^{1}$ }

\def\runauthor{ Gen {\sc SHIRANE} and Peter {\sc M.\ GEHRING} }

\inst{ Department of Physics, Brookhaven National Laboratory, Upton,
  New York 11973-5000, $^{1}$NIST Center for Neutron Research,
  National Institute of Standards and Technology, Gaithersburg,
  Maryland 20899-8562 }

\abst{ A review is given of the phonon anomalies, which have been
  termed ``waterfalls,'' that were recently discovered through a
  series of neutron inelastic scattering measurements on the
  lead-oxide relaxor systems PZN-$x$PT, PMN, and PZN.  We discuss a
  simple coupled-mode model that has been used successfully to
  describe the basic features of the waterfall, and which relates this
  unusual feature to the presence of polar micro-regions. }

\kword{relaxor, ferroelectric, soft mode, neutron scattering, PMN,
  PZN, PZN-$x$PT, polar micro-regions}

\begin{document}
\sloppy
\maketitle

\section{Introduction}

Our current phonon studies of relaxor ferroelectrics are part of a
systematic investigation of perovskite oxides that have exceptionally
high piezo responses.  Two solid solutions, the phase diagrams for
which are shown in Fig.~1, have been extensively investigated in
recent years.  A common feature of these two systems is the
morphotropic phase boundary (MPB) which separates the tetragonal and
rhombohedral ferroelectric phases.  This boundary, particularly for
Pb(Zr$_{1-x}$Ti$_x$)O$_3$ (PZT), is nearly vertical against the
concentration parameter $x$.  In both cases the maximum piezo activity
is located on the rhombohedral side of the MPB.  A key difference
between these two systems is the relaxor behavior of
Pb[(Zn$_{1/3}$Nb$_{2/3}$)$_{1-x}$Ti$_x$]O$_3$ (PZN-$x$PT), for which
the B-site of the end member PZN is occupied by the heterovalent ions
Zn$^{2+}$ and Nb$^{5+}$.  The mixed-valence character of the B-site
produces unique relaxor properties such as dielectric relaxation, and
the appearance of nanometer-sized polar domains in the cubic phase.
Extensive x-ray scattering experiments have already been carried out
to investigate the nature of the phase transitions near the MPB.  Here
we will review the current activity of neutron scattering studies of
soft phonons in relaxor ferroelectrics.

The recent discovery of a monoclinic phase by Noheda {\it et al.\ }
that intervenes between the rhombohedral and tetragonal phases (shown
as the hatched region in Fig.~1) has suggested a new interpretation of
the origin of the high piezoelectricity in PZT.~\cite{Noheda} This
result lead to the further experimental study of poled ceramics by Guo
{\it et al}.~\cite{Guo} A subsequent first-principles calculation has
given theoretical backing for the high piezo effect in the monoclinic
phase which permits additional freedom for the polar ions that is not
otherwise allowed in the other phases.~\cite{Bellaiche} The next stage
of this investigation was to have been a dynamical study of this
material, namely phonons.  Unfortunately, however, large single
crystals of PZT with compositions near the MPB are not available for
phonon studies.  Indeed, most work to date on this system has been
limited to ceramic samples.

%
%
\begin{figure}[h]
 \begin{center}
     \epsfxsize=11cm
     \epsfysize=14cm
     \epsfbox{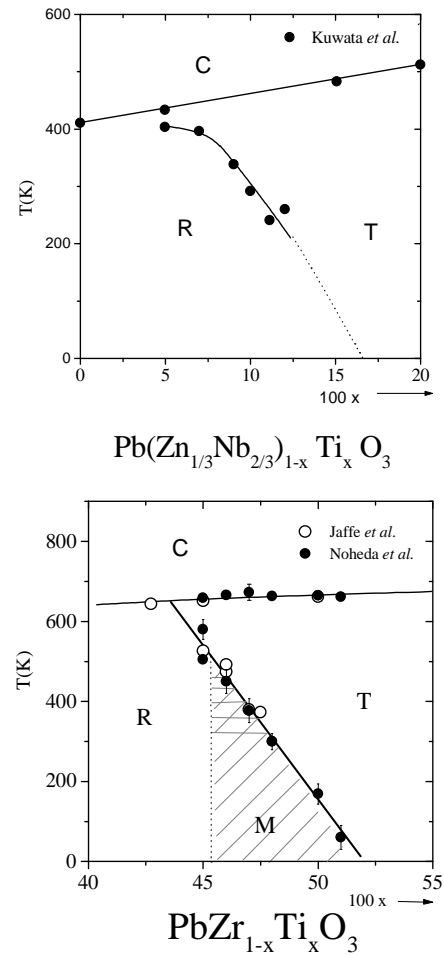}
 \end{center}
 \caption{ Phase diagrams for the PZN-PT and PZT systems.  The MPB is 
   represented by the solid line between the rhombohedral and
   tetragonal phases.  An intervening monoclinic phase has been
   discovered in PZT,~\cite{Noheda} and is speculated for PZN-PT. }
 \label{fig:1}
\end{figure}

On the other hand, Kuwata {\it et al.} produced, in 1982, large single
crystals of the relaxor ferroelectric PZN-$x$PT.~\cite{Kuwata} Neutron
inelastic scattering studies on similar single crystals has resulted
in the observation of a unique and unexpected phonon anomaly, the
so-called ``waterfall,'' by Gehring {\it et al.\ } for $x=0.08$
(PZN-8\%PT) which is shown in Fig.~2.~\cite{Gehring1} In this case,
the polar transverse optic (TO) phonons appear to drop precipitously
into the acoustic branch at a finite value of the reduced momentum
transfer $q=0.2$~\AA$^{-1}$, measured from the zone center, thereby
resembling a waterfall when plotted as a standard dispersion diagram.
It was speculated that this behavior is the result of the
nanometer-sized polar regions, also known as polar micro-regions
(PMR),~\cite{Tsurumi} that develop at temperatures far above $T_c$, a
phenomenon first proposed by Burns and Dacol.~\cite{Burns}

%
%
\begin{figure}[h]
 \begin{center}
     \epsfxsize=7cm
     \epsfysize=11cm
     \epsfbox{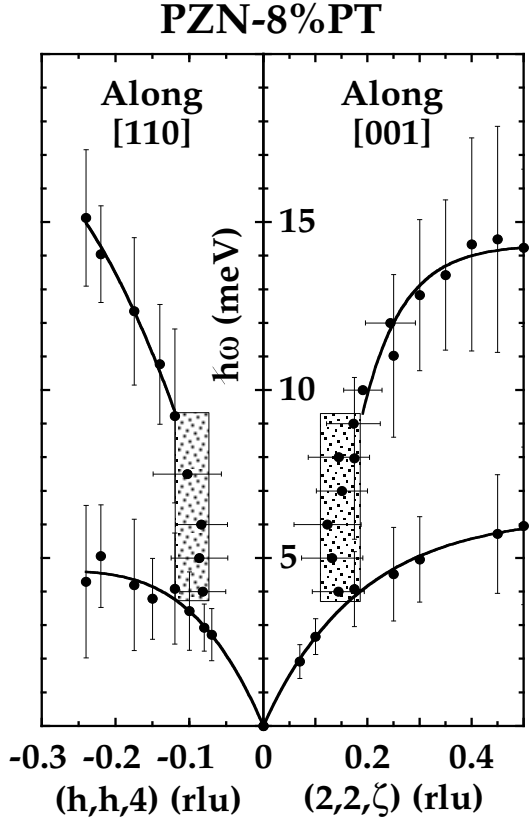}
 \end{center}
 \caption{ Solid dots represent positions of peak scattered neutron
   intensity taken from constant-$\vec{Q}$ and constant-$E$ scans at
   500~K along both [110] and [001] symmetry directions.  Vertical
   (horizontal) bars represent phonon FWHM linewidths in $\hbar
   \omega$ ($q$).  Solid lines are guides to the eye indicating the TA
   and TO phonon dispersions.  (From Ref.~[5].) }
 \label{fig:2}
\end{figure}

In this review article, we compare previous phonon measurements in PMN
(Pb(Mg$_{1/3}$Nb$_{2/3}$)O$_3$)~\cite{Naberezhnov,Gehring2} and
PZN-8\%PT,~\cite{Gehring1} as well as current work on pure
PZN.~\cite{Gehring3}  Very recently, a simple coupled-mode model was
proposed to relate the waterfall to the PMR.~\cite{Gehring3} We begin
with a survey of the relaxor properties of PMN, perhaps the most
typical and well-studied of the relaxor ferroelectrics.

\section{ PMN - Pb(Mg$_{1/3}$Nb$_{2/3}$)O$_3$ }

PMN is considered the prototypical relaxor compound, and has been
studied using a wide variety of different experimental techniques.
The real part of the dielectric susceptibility $\epsilon'$ exhibits a
broad peak at a temperature $T_{max}$ = 230~K that shifts to higher
temperature with increasing frequency.  The crystal structure remains
cubic down to 10~K;~\cite{Thomas} it exhibits a polar ferroelectric
phase around 212~K only when cooled under electric
field.~\cite{Westphal} In this respect PMN is very different from PZN
which undergoes a cubic -- rhombohedral phase transition around 410~K,
as shown in Fig.~1.  In 1983 Burns and Dacol proposed a seminal model
of the disorder intrinsic to relaxors.~\cite{Burns} Using measurements
of the optic index of refraction, they demonstrated that a
randomly-oriented local polarization $P_d$ develops in the form of
polar micro-regions (PMR), at a well defined temperature $T_d$.  This
temperature, often referred to as the Burns temperature, is about
610~K for PMN,~\cite{Burns} and is much higher than $T_{max}$ = 230~K.

%
%
\begin{figure}[h]
 \begin{center}
     \epsfxsize=7cm
     \epsfysize=15cm
     \epsfbox{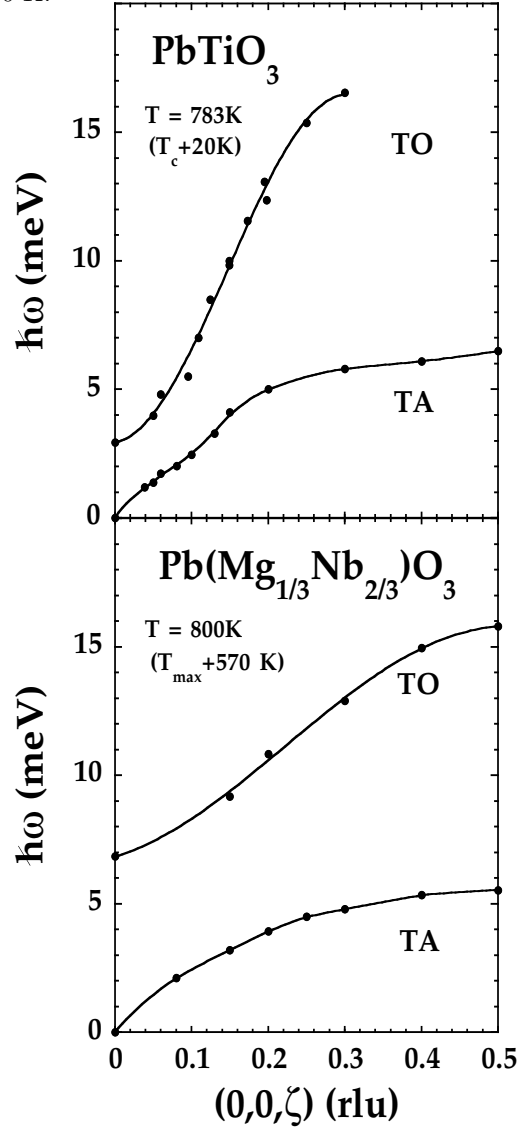}
 \end{center}
 \caption{    Top - Dispersion of the lowest energy TO mode and the 
   TA mode in PbTiO$_3$, measured just above $T_c$ (from Ref.~[13]).
   Bottom - Dispersion curves of the equivalent modes in PMN measured
   far above $T_{max}$ (from Ref.~[8]). }
 \label{fig:3}
\end{figure}

An extensive study of the low-frequency ($<$ 16~meV) phonons in PMN
was first reported by Naberezhnov {\it et al}.~\cite{Naberezhnov} They
covered a wide temperature range from 500~K to 900~K, thereby spanning
both sides of the Burns temperature $T_d$ = 610~K.  Data were
collected for small $q$ where the optic mode is overdamped, and the
analysis was done assuming a coupled-mode description.  Their phonon
dispersion curves measured at 800~K are shown in Fig.~3 together with
analogous data for PbTiO$_3$.~\cite{Shirane} The transverse optic mode
TO (referred to as TO1) appears to be the expected soft mode branch
common to all other ferroelectric oxides.  However, the assumption was
made for PMN that the TO1 branch could not be the same soft mode
branch because the $Q$-dependence of the structure factor was
inconsistent with that expected for the ferroelectric fluctuations.
This assumption does not appear to be justified as we will demonstrate
later. (Naberezhnov {\it et al.} introduced a fictitious QO branch
that was derived solely from the overdamped mode in the context of
their mode-coupling analysis.)  The series of neutron measurements
summarized in this review demonstrate that the normal TO phonon branch
of PMN, shown in Fig.~3, actually transforms into the waterfall below
the Burns temperature.

%
%
\begin{figure}[h]
 \begin{center}
     \epsfxsize=7cm
     \epsfysize=9cm
     \epsfbox{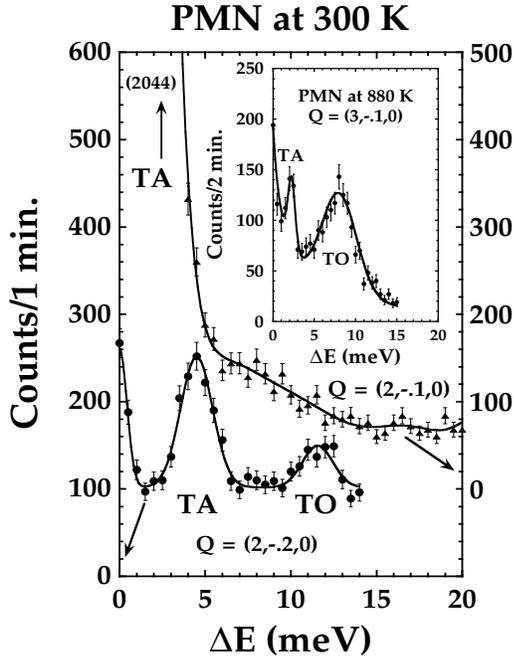}
 \end{center}
 \caption{Data from constant-$\vec{Q}$ scans taken near (200) at
   300~K.  Lines are guides to the eye.  The scan at $q = -0.2$ rlu
   shows well-defined TA and TO modes.  But at $q = -0.1$ rlu, only
   the TA peak is well-defined.  The TO mode is strongly overdamped.
   The inset, however, shows data taken on the same crystal at the
   same $q$ at 880~K in which the TO mode is clearly well-defined.
   (From Ref.~[9].) }
 \label{fig:4}
\end{figure}

A separate phonon study of PMN was carried out at room temperature at
the NIST Center for Neutron Research in 1997, and some of the results
are shown in Fig.~4.~\cite{Gehring2} These data were collected prior
to the discovery of the waterfall in PZN-8\%PT shown in Fig.~2.  In
marked contrast to the normal constant-$\vec{Q}$ scan in PMN observed
by Naberezhnov {\it et al.} at 880~K, shown in the inset to Fig.~4,
for which well-defined peaks are clearly visible for both the TA and
TO modes, the TO phonon cross section for $q$ = -0.10~rlu is
mysteriously absent at 300~K.  Only after the discovery of the
waterfall in PZN-8\%PT was it realized that the same anomaly was
present in PMN as well, and was the cause of the missing TO branch at
low $q$ and low temperature.  PMN, it should be mentioned, possesses
the advantage of having a relatively low Burns temperature compared to
that of other relaxors.  Hence the optic modes in PMN can be studied
above $T_d$, in the absence of any PMR.  The PZN system, on the other
hand, decomposes near its (higher) Burns temperature.

\section{Model Description of the Waterfall}

Phonon measurements on relaxors have since been extended to pure PZN,
the $x$=0 end member compound shown in the PZN-$x$PT phase diagram in
Fig.~1. During the course of these experiments, a simple but effective
model for the anomalous waterfall was developed.  We will describe
this model now before we present the new results from the detailed
study of the phonon cross sections in PZN.~\cite{Gehring3} In the case
of neutron energy loss, the scattering intensity distribution $I$ for
two interacting modes with frequencies $\Omega_1$ and $\Omega_2$, and
widths $\Gamma_1$ and $\Gamma_2$, is given by the
expression~\cite{Bullock}

\begin{eqnarray}
I &\sim &[n(\omega) + 1]  \frac{\omega}{A^2 + \omega^2 B^2} \times \nonumber \\ 
  &     &( [( \Omega_2^2 - \omega^2 )B - \Gamma_2 A]F_1^2 + 2\lambda BF_1 F_2 + \nonumber \\
  &     &  [( \Omega_1^2 - \omega^2 )B - \Gamma_1 A]F_2^2 ),
\end{eqnarray}

\noindent
where $A$ and $B$ are given by

\begin{eqnarray}
A &= &(\Omega_1^2 - \omega^2)(\Omega_2^2 - \omega^2) - \omega^2\Gamma_1\Gamma_2, \nonumber \\
B &= &\Gamma_1(\Omega_2^2 - \omega^2) + \Gamma_2(\Omega_1^2 - \omega^2),
\end{eqnarray}

\noindent 
and $n(\omega)$ is simply the Bose factor $[e^{(\omega/k_BT)} -
1]^{-1}$.  The quantities $F_{1,2}$ are the structure factors of modes
1 and 2, and $\lambda$ is the coupling strength between the two modes.
This equation has been shown to describe the behavior of coupled-phonon
cross sections quite well.~\cite{Gehring3,Harada,Bullock}

The essential physics behind the mode-coupled description of the
low-frequency dynamics of relaxor ferroelectrics is built into the
linewidth of the optic mode $\Gamma_1$, which is assumed to become
sharply $q$-dependent when the polar micro-regions are formed at the
Burn's temperature $T_d$.  If we suppose that the PMR have an average
diameter given by $2\pi/q_{wf}$, where $q_{wf}$ represents the
reciprocal space position of the waterfall, then those optic phonons
having $q < q_{wf}$ will not be able to propagate easily because their
wavelength exceeds the average size of the PMR.  These polar lattice
vibrations are effectively impeded by the boundary of the PMR.  The
simplest way to simulate this situation is to assume a sudden and
steep increase in $\Gamma_1$ at $q_{wf}$.  Fig.~5 shows several model
constant-$\vec{Q}$ simulations based on this assumption using the
value $q_{wf}$ = 0.15~rlu.~\cite{Gehring3,Ohwada} For simplicity, the
dispersions of both optic and acoustic modes were ignored by holding
the parameters $\Omega_{1,2}$ fixed at 10 and 5~meV, respectively,
over the entire Brillouin zone.

%
%
\begin{figure}[h]
 \begin{center}
     \epsfxsize=7cm
     \epsfysize=11cm
     \epsfbox{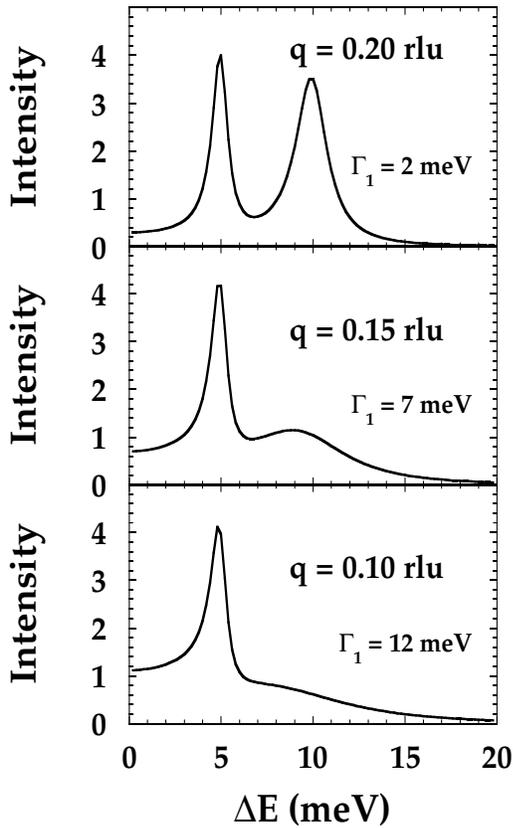}
 \end{center}
 \caption{Model simulations assuming a coupled-mode intensity distribution
   and a strongly $q$-dependent TO phonon linewidth
   $\Gamma_1$.~\cite{Gehring3,Ohwada} Three constant-$\vec{Q}$ scans
   are shown corresponding to $q$ = 0.20, 0.15, and 0.10~rlu, with $q$
   = 0.15~rlu representing the reciprocal space position of the
   anomalous waterfall feature. }
 \label{fig:5}
\end{figure}

For $q > q_{wf}$ one observes two broad peaks, as expected.  At
momentum transfers $q$ below $q_{wf}$, however, the optic mode becomes
highly overdamped and its profile extends in energy below that of the
acoustic mode.  Alongside each constant-$\vec{Q}$ scan is shown the
corresponding value of $\Gamma_1$ used in the simulation.  The
waterfall thus represents the crossover between a high-$q$ regime, in
which one observes two well-defined peaks corresponding to two
propagating modes, and a low-$q$ regime, in which one observes an
overdamped optic mode plus an acoustic peak.  This simple model cross
section describes all of the experimental observations very well.
Indeed, one can favorably compare the simulated scan at $q$ = 0.10~rlu
in Fig.~5 with the corresponding experimental scan shown in Fig.~4.
One can see now that the waterfall is not actually an enhancement of
the phonon scattering cross section.  Instead, it is simply a
redistribution of the optic mode profile that is caused by the polar
micro-regions which induce a sudden change in the optic mode linewidth
at a specific $q$ that is related to the average size of the PMR.  In
order to illustrate the basic characteristics of the coupled-mode
model scattering cross section, the $q$-dependence of the optic mode
linewidth $\Gamma_1$ is shown in Fig.~6 along with two simulated
constant-$E$ scans at 0 and 7.5~meV.  It is apparent that the sharp
increase in $\Gamma_1$ has a pronounced effect on both cross sections
in the vicinity of $q_{wf}$.

%
%
\begin{figure}[h]
 \begin{center}
     \epsfxsize=7cm
     \epsfysize=11cm
     \epsfbox{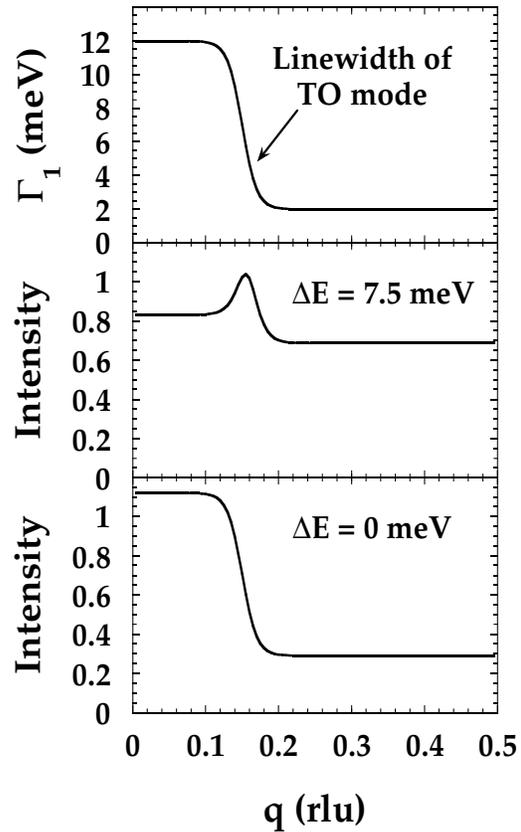}
 \end{center}
 \caption{ Model simulations of two constant-$E$ scans at 0 and 
   7.5~meV.~\cite{Gehring3,Ohwada} The $q$-dependence of the TO
   linewidth $\Gamma_1$ is shown in the top panel for ease of
   comparison. }
 \label{fig:6}
\end{figure}

\section{PZN - Pb(Zn$_{1/3}$Nb$_{2/3}$)O$_3$}

Extensive surveys of the phonon profiles in pure PZN have now been
completed. Two large, high quality single crystals, weighing 4.4 and
2.3 grams, were examined and gave identical results.  These crystals
were mounted in the $(hk0)$ zone, and the data were taken on the BT2
triple-axis spectrometer located at the NIST Center for Neutron
Research.  At low temperatures a neutron energy loss configuration was
used with a fixed final energy of 14.7~meV, whereas a neutron energy
gain configuration with a fixed incident energy of 14.7~meV was
employed at high temperatures.  PZN has a cubic-to-rhombohedral phase
transition around 410~K, so data were collected at 500~K, well above
the ferroelectric transition.  Fig.~7 shows several typical
constant-$E$ scans at 500~K for the 4.4 gram PZN crystal.  Both the -6
and -8~meV (negative for neutron energy gain) scans are centered at
the same $q$, namely $q_{wf}$, indicating the presence of the
waterfall anomaly in PZN.  These two scans are to be compared to the
7.5~meV scan shown in Fig.~6.  At a higher energy transfer of -12~meV,
the scattering has shifted to higher $q > q_{wf}$, and one recovers a
normal propagating TO mode.

%
%
\begin{figure}[h]
 \begin{center}
     \epsfxsize=7cm
     \epsfysize=10cm
     \epsfbox{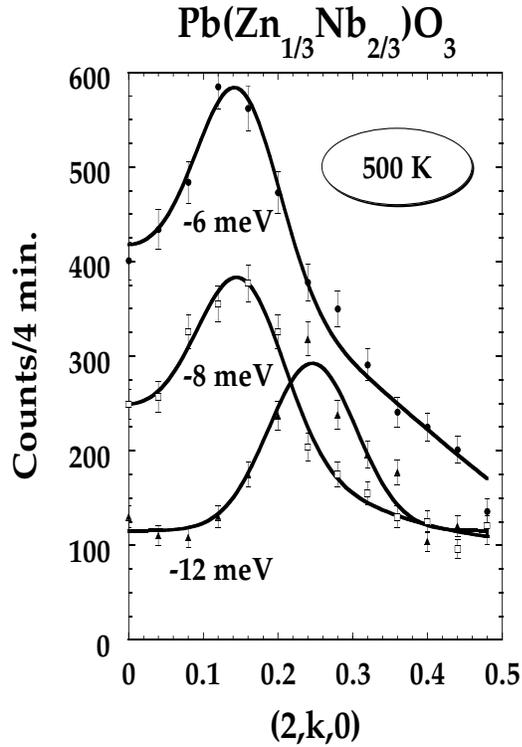}
 \end{center}
 \caption{ Three constant-$E$ scans taken on PZN at 500~K using neutron 
   energy gain. The scans at -6 and -8~meV are centered at the same
   $q$, indicating the presence of the waterfall.  The -12~meV scan is
   shifted to higher $q$ and lies outside the waterfall region. (From
   Ref.~[10].)}
 \label{fig:7}
\end{figure}

In order to produce a detailed contour map of the scattering intensity
in the cubic phase, constant-$\vec{Q}$ scans were taken in 0.04~rlu
steps across the entire Brillouin zone near (200).  The results are
shown in Fig.~8.  A constant background has been subtracted from the
data, and the resulting intensities placed on a logarithmic color
scale.  Both the TO and TA phonon branches are clearly visible above
$q_{wf} \sim 0.14$~rlu.  At $q_{wf}$, however, the waterfall is
readily visible as the broad vertical red feature that appears to drop
into the TA phonon branch.  The TA mode appears saturated (yellow
color) below 5~meV only because the intensities displayed in Fig.~8
were limited in their range to better show the waterfall.

%
%
\begin{figure}[h]
 \begin{center}
   \epsfxsize=7cm 
   \epsfysize=10cm 
   \epsfbox{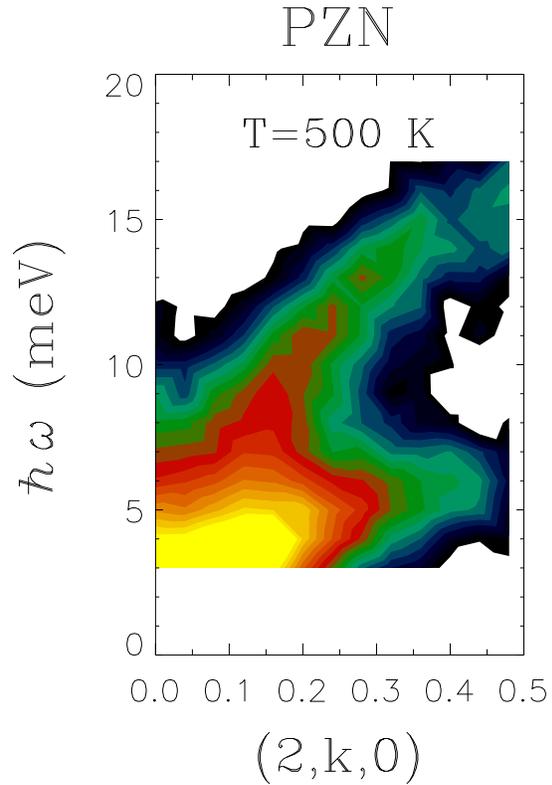}
 \end{center}
 \caption{ Contour map of the background subtracted scattering intensity 
   from PZN at 500~K measured near (200).  The intensity is indicated
   by a logarithmic color scale that is limited to a narrow range in
   order to better show the waterfall.  Yellow is the most intense.
   (From Ref.~[10].)}
 \label{fig:8}
\end{figure}

An attempt to measure the high-temperature ($T > T_d$) limiting
behavior of phonons for $q < q_{wf}$ was made using
PZN-8\%PT.~\cite{Gehring1} Unfortunately, due to the high Burns
temperature for PZN, it was not possible to reach sufficiently high
temperatures to recover a normal TO phonon dispersion; at present PMN
appears to offer the best chance to study this aspect of the waterfall
problem.  In the case of PZN, the evolution of the phonons through the
phase transition at 410~K and to lower temperatures have been
investigated.~\cite{Gehring3} The intensity of the anomalous waterfall
decreases gradually through the phase transition and almost disappears
at 150~K.  At the same time a normal optic mode near $q$=0 is
recovered at 11~meV.  At present this result is interpreted to mean
that the polar micro-regions have become sufficiently large at low
temperatures to permit the propagation of long-wavelength optic modes.

Our current experimental results give a good estimate of the linewidth
of the optical branch as a function of $q$.  The simplest model
identifies the $q$ for the waterfall with the average size of the PMR.
A more sophisticated theoretical model is clearly needed to relate the
distribution of sizes of the PMR to the observed phonon profiles.
Currently we are examing the effects of an applied electric-field on
the waterfall and low-$q$ phonons to determine the effect of the field
on the PMR.

\section*{Acknowledgements}

We thank Y.\ Fujii, K.\ Hirota, B.\ Noheda, K.\ Ohwada, S.\ Park, S.\ 
Vakrushev, and H.\ You for stimulating discussions.  Financial support
by the U.\ S.\ Dept.\ of Energy under contract No.\ DE-AC02-98CH10886
is acknowledged.  We also acknowledge the support of the NIST Center
for Neutron Research, U.\ S.\ Dept.\ of Commerce, in providing the
neutron facilities used in this work.

\end{document}